# Properties of Binary Transition-Metal Arsenides (*T*As)

Bayrammurad Saparov, Jonathan E. Mitchell, Athena S. Sefat

*Materials Science and Technology Division, Oak Ridge National Laboratory, Oak Ridge, Tennessee 37831-6114, USA*

**Abstract**

We present thermodynamic and transport properties of transition-metal (*T*) arsenides, *T*As with *T* = Sc to Ni (3*d*), Zr, Nb, Ru (4*d*), Hf and Ta (5*d*). Characterization of these binaries is made with powder X-ray diffraction, temperature and field-dependent magnetization and resistivity, temperature-dependent heat capacity, Seebeck coefficient, and thermal conductivity. All binaries show metallic behavior except TaAs and RuAs. TaAs, NbAs, ScAs and ZrAs are diamagnetic, while CoAs, VAs, TiAs, NiAs and RuAs show approximately Pauli paramagnetic behavior. FeAs and CrAs undergo antiferromagnetic order below $T_N \approx 71$ K and $T_N \approx 260$ K, respectively. MnAs is a ferromagnet below $T_C \approx 317$ K and undergoes hexagonal-orthorhombic-hexagonal transitions at $T_S \approx 317$ K and 384 K, respectively. For *T*As, Seebeck coefficients vary between + 40 μV/K and - 40 μV/K in the 2 K to 300 K range, whereas thermal conductivity values stay below 18 W/(m K). The Sommerfeld-coefficient $\gamma$ are less than 10 mJ/($K^2$mol). At room temperature with application of 8 Tesla magnetic field, large positive magnetoresistance is found for TaAs (~25%), MnAs (~90%) and for NbAs (~75%).

## 1. Introduction

Since the discovery of superconductivity in an iron-containing material, $LaFeAsO_{1-x}F_x$ ($x \approx 0.11$) with $T_C = 26$ K [1], ongoing research has been devoted to finding other iron-based superconductors (FeSCs) [2]. The $T_C$ values reach as high as ≈ 55 K for the so-called '1111' quaternary materials with tetragonal ZrCuSiAs-type structure in $Gd_{0.8}Th_{0.2}FeAsO$, $SmFeAsO_{0.9}F_{0.1}$, $SmFeAsO_{0.85}$, and $SmFeAsO_{0.8}H_{0.2}$ compounds [3-6]. These materials show competition between superconductivity and magnetism with the application of pressure or chemical doping [2, 7]. Their quasi-two-dimensional structures feature FeAs covalent layers in the *ab*-plane that are separated by rare-earth oxide layers along the *c*-axis [2]. There is a continuing demand for FeSCs to reach high $T_C$ values; therefore the fundamental factors that can achieve this effect are of high interest. While FeAs layers are found to be crucial [2], high tetrahedral symmetry (~109.5°) [8], arsenic atomic height from the Fe-plane of $z$ =1.38 Å [9, 10], and a large FeAs interlayer spacing [11] may be important as well. There are numerous ternary transition-metal-based arsenides with similar tetragonal structures to those of FeSCs, but the non-Fe containing materials have given small $T_C$ values at best [12-14]. Binary transition-metal arsenides (*T*As) can be considered as 'proxy' structures of potential arsenide-based superconductors in the same way that CuO is a proxy structure [15] of cuprate superconductors. In this manuscript, we report synthesis, thermodynamic and transport properties of *T*As with *T* = Sc to Ni (3*d*), Zr, Nb, Ru (4*d*), Hf and Ta (5*d*) and report trends across the 3*d*, 4*d*, and 5*d* transition-metal series. The only *T*As binaries omitted from this study are those of RhAs (expensive), HgAs (toxic), and YAs (arguably rare-earth based). In addition to serving as proxy materials, the properties of *T*As can serve to identify the effects of binary arsenide impurities caused by transition-metal doping of FeSC.

The *T*As crystal structures bear two differences compared with 1111 structures of FeSCs (Fig. 1). First, their structures are made of 3D networks, contrasting quasi-layers in the FeSCs. Second, they are built from transition-metal-centered $TAs_6$ octahedra (Fig. 1b, 1d-f), or trigonal prisms (Fig. 1c), unlike 1111 that has *T* tetrahedral ($TAs_4$) geometry (Fig. 1a). They crystallize in five structure types: face-centered cubic NaCl-type (for ScAs); hexagonal TiP-type (for TiAs and ZrAs, HfAs); orthorhombic MnP-type (for VAs, CrAs, FeAs, CoAs and RuAs); hexagonal NiAs-type (for MnAs and NiAs); tetragonal NbAs-type structure (for NbAs and TaAs). The room-temperature lattice parameters from literature [16-26] are summarized in Table 1. The variation of the structure types is similar to that noted for transition metal phosphides. The structural transitions in phosphides have been attributed to electronic effect, and in particular, to a second order Jahn-Teller distortion [27].

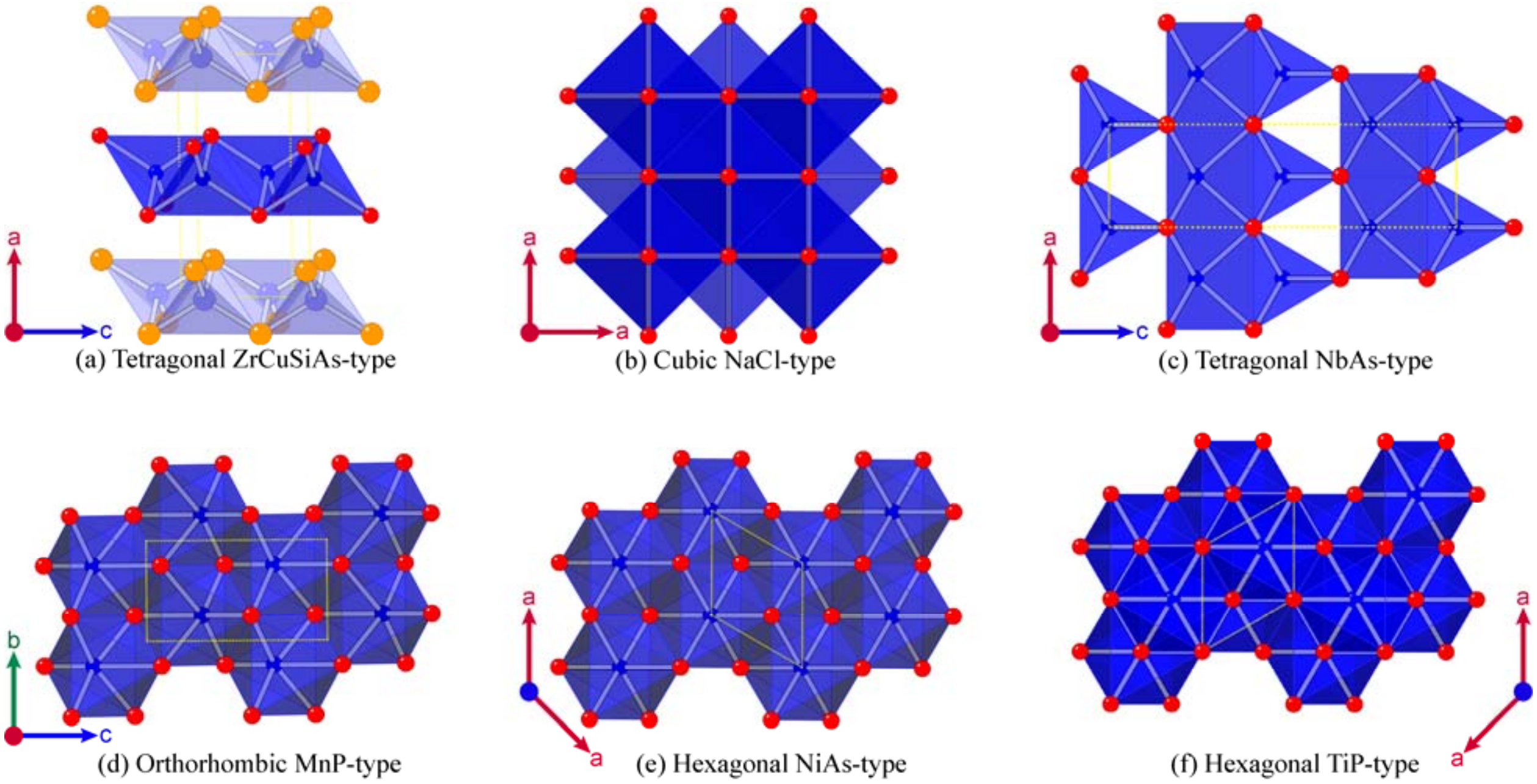

**Figure 1**: Crystal structures of (a) tetragonal ZrCuSiAs-type (1111 FeSC), (b) cubic NaCl-type (ScAs) (c) tetragonal NbAs-type (NbAs and TaAs), (d) orthorhombic MnP-type (VAs, CrAs, FeAs, CoAs and RuAs), (e) hexagonal NiAs-type (MnAs and NiAs) and (f) hexagonal TiP-type (TiAs, ZrAs and HfAs). Transition metals are shown in red, and arsenic atoms are shown in blue. The coordination of *T* is tetrahedral in ZrCuSiAs-type, octahedral in NaCl-, MnP- and NiAs-type structures, and trigonal-prismatic for NbAs-type.

There is a small number of experimental studies on transition-metal arsenide binaries, mainly focused on 3*d* materials that reveal variation of magnetic and electronic properties across *T* series. ScAs is reported to exhibit a negative temperature coefficient of the electrical resistivity, indicating semiconducting behavior with $\rho_{295K}$= 12.6×$10^3$ mΩ.cm and Seebeck coefficient of $S_{295K}$ = -70 μV/°C [28]. VAs is Pauli paramagnetic [17, 29]. For CrAs, there is a transition upon increasing temperature at $T_{N,i}$ ≈ 270 K that is related to the double-spiral $H_c$-type antiferromagnetic transition [30-32]. The spin-lattice coupling causes this transition to be first order, with discontinuities in the lattice constants and a thermal hysteresis ($T_{N,d}$ ≈ 260 K with decreasing temperature) [17, 33-34]. In the ordered state, neutron diffraction determines the magnetic moment of 1.70(5) $\mu_B$/Cr [17]. The anomaly in the heat capacity gives entropy of $S$ = 0.69 cal/Kmol [35]. There is a second transition at $T_D$ ≈ 1170 K, related to the structural distortion from the low temperature orthorhombic MnP- to hexagonal NiAs-type structure ($S$ = 0.22 cal/Kmol) that involves a continuous shift of Cr and As atoms [35, 36]. MnAs is a ferromagnet with a first-order transition at Curie temperature $T_C$ ≈ 316 K, accompanied by a structural transformation into an orthorhombic MnP-type structure in a paramagnetic state. The transition temperature and thermal hysteresis are strongly dependent on the applied field ($T_C$ = 335 K at 6 T; ΔT = 5 to 10 K) [21, 37]. MnP-type structure reverts to hexagonal NiAs-type at $T_S$ ≈ 398 K in a second order transition which involves distortion of $TAs_6$ octahedra [38, 39]. A large magnetocaloric effect is produced, with a magnetic entropy of $S$ ≈ 4 to 5 J/(K mol) [40, 41]. The saturation magnetization is 3.4 $\mu_B$/Mn [43]. For FeAs, antiferromagnetic order is observed below $T_N$ ≈ 70 K [22, 43, 44]. The incommensurate magnetic structure was initially described as helimagnetic having double spiral arrangement in the *c*-axis spiral direction [21], but recently there is evidence for a non-collinear spin-density-wave structure [45]. The magnetic susceptibility data fitted Curie-Weiss law between ~ 300 K and 650 K, giving $\theta$ ≈ -1400 K and 3.1 $\mu_B$/Fe [21]. For CoAs, no long range magnetic order was observed, although the magnetic susceptibility results show a broad minimum at ~ 225 K, with Curie-Weiss behavior above 490 K giving $\theta$ ≈ -230 K and 1.1 $\mu_B$/Co [46]. It is also noted that CoAs is a metal [46], however, the measurement data is not provided in literature. There is very little information on 4*d* and 5*d* *T*As binaries. RuAs is Pauli paramagnetic and has poor metallic behavior with metal-insulator transition at ≈ 200 K [47]. The broad peak around 280 K in the magnetic susceptibility data for RuAs has been ascribed to a possible phase transition [47]. NbAs and TaAs have diamagnetic behavior [48, 49].

This manuscript reports on X-ray diffraction, temperature and field-dependent magnetization and resistivity, temperature-dependent heat capacity, Seebeck coefficient, and thermal conductivity for *T*As polycrystalline samples of *T* = Sc to Ni (3*d*), Zr, Nb, Ru (4*d*), Hf and Ta (5*d*). Although there is a collection of isolated studies on a few of the *T*As binaries as described above, they are mainly old publications which usually do not report temperature-dependence of thermodynamic and transport data. The resistivity data for TiAs, VAs, MnAs, CoAs, NiAs, ZrAs, NbAs and TaAs are reported here for the first time. In addition, thermal conductivity and Seebeck coefficient data for *T*As are also original, with the exception that data on MnAs [50] and ScAs [24] were reported before. We believe that our study provides a useful reference for the condensed matter physics community since it presents the magnetic and electronic properties across the transition-metal series. Moreover, it may provide clues as to the importance of structure type for the occurrence of superconductivity in FeSCs. Superconductivity may also be induced by doping the binary materials themselves, as demonstrated recently with the substitution of 25% Rh in RuAs, producing $T_C \approx 2$ K [47]. Finally our results may function as a useful reference for recognizing the effects of *T*As binary impurities in chemical substitution studies or reports on impure samples [51, 52]. Below we present a brief review of synthesis and structure, which is then followed by the presentation of the survey of thermodynamic and transport properties for each material.

## 2. Experimental details

The binaries were synthesized using metals (powder, foil or small pieces) and arsenic pieces from Alfa Aesar or Ames Laboratory, which had purities greater than 99.9%. The stoichiometric molar ratio of *T*:As (*T* = Sc, Ti, V, Cr, Mn, Fe, Co, Ni, Zr, Nb, Ru, Hf, and Ta) were each loaded into a silica ampoule and sealed under vacuum. Reaction mixtures were initially heated to 600 ºC or 700 ºC in a box furnace, annealed for up to 10 hrs, then gradually heated (30 ºC/hr) to the final sintering temperature (650 ºC for CoAs; 900 ºC for ScAs, VAs, MnAs, CrAs and NiAs; 1000 ºC for TiAs, ZrAs, NbAs, RuAs, HfAs, TaAs, and re-sintered ScAs; 1065 ºC for FeAs), held for 10 to 24 hours, then furnace cooled. In order to improve phase purity, a single regrinding and reheating step was required for TiAs, CrAs, and MnAs, NbAs, HfAs, and TaAs, while ScAs and VAs required at least five re-sintering steps. The resulting materials were uniaxially pressed into pellets and annealed in an evacuated silica ampoule at 900 or 1000 °C for 24 to 48 hours. Those of poor consistency were reground, re-pelletized and re-annealed. The pellet densities were material dependent and measured in the range of 57 to 82% of the X-ray diffraction theoretical values.

Small chunks of the pellets were ground and examined using a PANalytical X'Pert PRO MPD x-ray powder diffractometer, using the Ni-filtered Cu-K$\alpha$ radiation and detector with 2.122° active length. Typical runs were collected in 5-90° 2$\theta$ range. Even though all samples were confirmed to be stable under ambient air for several hours by consecutive X-ray powder diffraction experiments, they were stored inside a helium-filled glove-box, as a precautionary measure. Data collections that were aimed at crystal structure refinements were carried out in 10-90° 2$\theta$ range, with a step size of 1/60° and a counting time of 120 seconds/step in a continuous scan mode. Lattice parameters were determined by Rietveld refinement using GSAS-EXPGUI [53] software package, results of which are summarized in Table 1; the literature values are also reported here. Impurity phases were estimated using the Hill and Howard method [54] and their total contents were less than 5% by mass, except for HfAs.

The measurements of magnetization, heat capacity, electrical resistivity, Seebeck coefficient, and thermal conductivity were carried out using pieces of polycrystalline pellets. Thermal transport measurements were not measured for MnAs because we were unable to prepare a large enough bar for contacts. In addition, we do not report electric and thermal transport properties for CrAs because the polycrystalline bars consistently broke on cooling. Moreover, we have omitted transport properties for HfAs because of poor sample quality.

Magnetization results were collected using a Quantum Design Magnetic Property Measurement System (MPMS). For a typical temperature-sweep experiment, M(T), the sample was cooled to 2

K in zero-field (zfc) and data were collected by warming from 2 K to ~ 380 K in 1 Tesla, then collected by cooling in field (fc). In addition, zfc magnetization experiments were carried out from 350 K to ≤ 750 K for selected binaries with the use of an oven insert. Field-dependent magnetization, M(H), was typically measured at 2 K and room temperature with increasing field from zero. The measurements of heat capacity, electrical resistivity, Seebeck coefficient, and thermal conductivity were performed on a Quantum Design Physical Property Measurement System (PPMS). The temperature dependence of heat capacity, $C_p(T)$, was measured below 200 K via the relaxation method. For transport measurements, pellets were cut into bars measuring ≈ 8 x 3 x 1 $mm^3$ using a diamond-edged circular saw. For temperature- and field-dependence of electrical resistivity, ρ(T), four platinum leads were attached to the sample using Dupont 4929 silver paste. Resistance was measured upon warming from 2 K to ~ 400 K, then measured upon cooling in a field of 8 Tesla. For temperature dependence of Seebeck coefficient, *S*(T), and thermal conductivity, κ(T), electrical and thermal connections to each sample were made using gold-plated copper leads with the aid of Epotek H20E silver epoxy. Typical measurements were done in the 2 K to 300 K range. With our PPMS setup, thermal conductivity values above 200 K are probably overestimated due to some heat lost through radiation.

## 2. Results and discussion

Table 1 contains the room-temperature crystallographic space groups for *T*As series whose structures were displayed in Figure 1. The cubic crystal structure for ScAs is unique in this series and may be rationalized if a complete charge-transfer from Sc to As is assumed according to $Sc^{3+}As^{3-}$, making it isoelectronic with NaCl. Both the TiP- and the MnP-type structure may be derived from the NiAs-type structure, with the TiP-type being a superstructure of the NiAs lattice with a doubled *c*-paramater, while in the MnP-type the metal atoms are shifted from the center of the $TAs_6$ octahedra, leading to the lower symmetry orthorhombic space group. NbAs crystal structure is different as it is made from $NbAs_6$ trigonal prisms that are stacked in 3D layers.

The refined lattice parameters for our synthesized *T*As samples are also summarized in Table 1; they compare well with the literature values [16-26]. Early reports of VAs [29], FeAs [55] and CoAs [46] structures were assigned to the $Pna2_1$ (No. 33) space group, which is a subgroup of *Pnma* (No. 62). This conclusion was based on the results of the Hamilton significance test [56] and the deviation from $y_{As}$ = 1/4, which is required for the 4*c* special position in the higher symmetry space group. The latest reports refined FeAs, CoAs and VAs in the space group *Pnma* [17, 57]. Here, the X-ray diffraction data did not converge when refining with $Pna2_1$, but could only be refined in *Pnma*, and hence, this is our chosen space group.

Several of the binaries contained minor impurity phases and their relative proportions by weight are summarized in Table 1. ScAs in particular contained several small peaks which could not be securely identified. Figure 2 is a Rietveld refinement of FeAs X-ray diffraction pattern, shown as a representative refinement for *T*As materials, giving very good reliability values of profile factor of $R_p$ = 0.0118 and weighted profile factor $R_{wp}$ = 0.0175. In fact, the quality of all data refinements were very good, resulting in a range of values of $R_p$ = 0.0112 to 0.0421 and $R_{wp}$ =

0.0157 to 0.0562 that are listed in Table 1. This is with the exception of HfAs, which had a high oxide content.

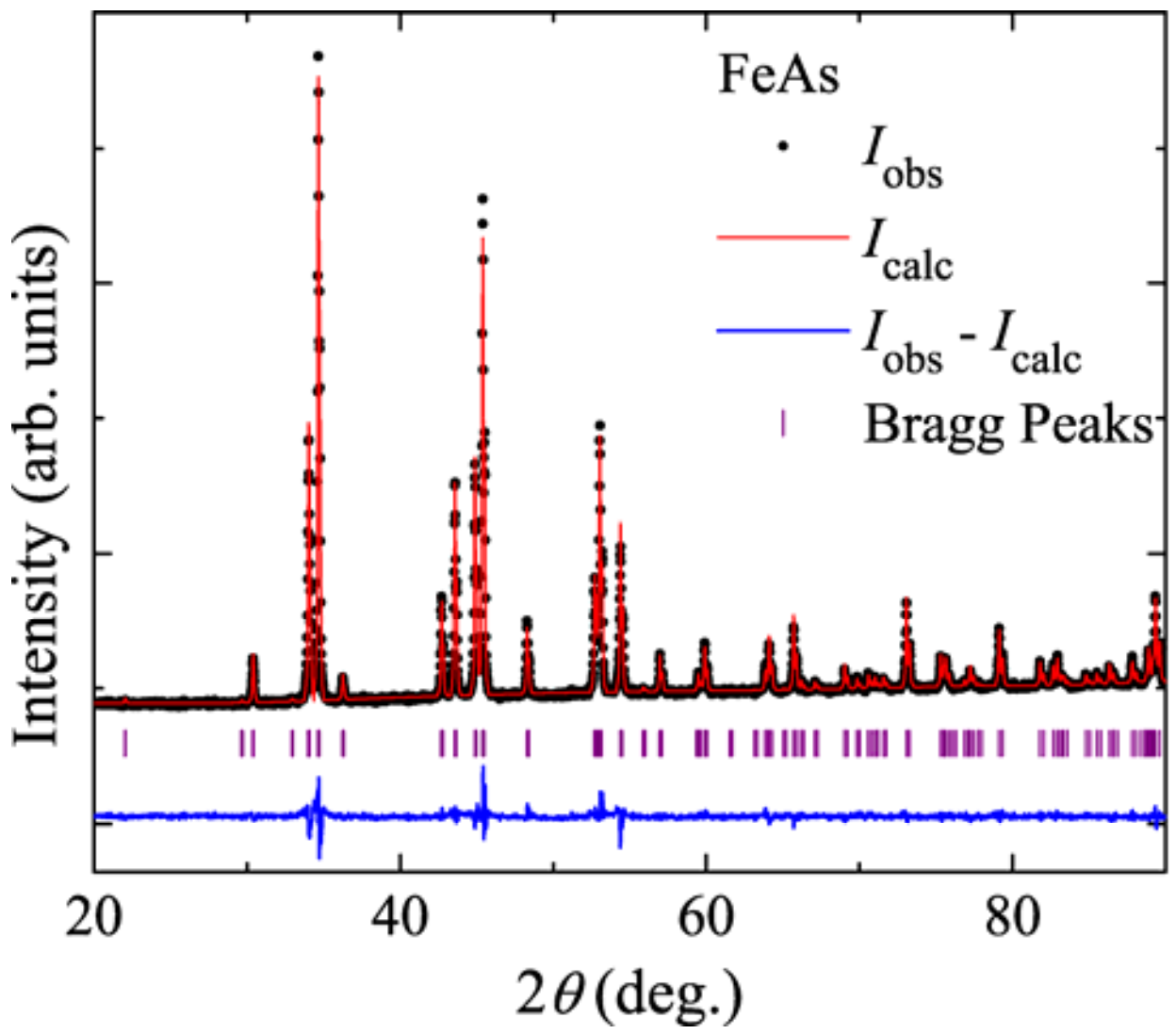

**Figure 2**: Powder X-ray diffraction pattern for FeAs (black dots), Rietveld fit (red line) to *Pnma* structure (Bragg positions in purple ticks), and the difference plot in blue.

Temperature-dependence of electrical resistivity results, ρ(T), are plotted in Figure 3, the first time published for $T$ = Ti, V, Mn, Co, Ni, Zr, Nb, and Ta. The room temperature values range from $\rho_{300K} \approx 0.03$ mΩ.cm for NiAs to $\rho_{300K} \approx 1.9$ mΩ.cm for MnAs (Table 2). Typical metallic behavior is observed for 3$d$-based $T$As with $T$ = Sc, Ti, V, Co, Ni (Fig. 3a, b) where $\rho$ increases with temperature in 2 K to 400 K range. For ScAs, metallic behavior was seen in electronic structure calculations [58] in support of our data. These findings are in contradiction to a semiconducting behavior expected by Zintl formalism [59] and reported experimentally [28].

For metallic $T$ = Mn, there is a discontinuity in $\rho$ at the first-order ferromagnetic-to-paramagnetic transition at $T_C \approx 316$ K that is accompanied by hexagonal-to-orthorhombic structural transition (Fig. 3d). For $T$ = Fe, $\rho$ increases with temperature, showing a small anomaly at the antiferromagnetic transition temperature of $T_N \approx 71$ K, then becoming approximately temperature independent above 200 K (Fig. 3d). Metallic behavior is shown for early 4$d$ $T$As with $T$ = Zr, Nb (Fig. 3a, b), whereas semiconducting behavior is observed for Ru and 5$d$ $T$ = Ta (Fig. 3c). RuAs shows temperature-independent resistivity above ~240 K similar to that reported [47]; a small band gap of ~ 0.01 eV can be derived from the Arrhenius fit ($\ln\rho = \ln\rho_0 - E_g/2k_B\mathrm{T}$) in the 110 K to 330 K region.

**Table 1**: For *T*As transition-metal arsenides, space group, refined unit-cell lattice-parameters, impurity phases, and refinement factors. Estimates of mass percent of impurity phases are given in parentheses. Literature lattice parameters and reference numbers are provided for comparison.

| Phase | Space Group | Refined Lattice Parameter (Å) | Impurities (%) | Refinement Parameter Values | | | Literature Lattice Parameter (Å) |
|---|---|---|---|---|---|---|---|
| | | | | $R_p$ | $R_{wp}$ | $\chi^2$ | |
| ScAs | *Fm*-3*m* | *a* = 5.46396(3) | Multiple (~4) | 0.0367 | 0.0545 | 2.61 | *a* = 5.4640(2) [16] |
| TiAs | *P*6$_3$/*mmc* | *a* = 3.64305(1), *c* = 12.05552(5) | $SiO_2$ (trace) | 0.0254 | 0.0362 | 1.73 | *a* = 3.6419(2), *c* = 12.055(1) [16] |
| VAs | *Pnma* | *a* = 5.85048(7), *b* = 3.36643(4), *c* = 6.28868(7) | $V_4As_3$ (1.7) | 0.0328 | 0.0461 | 2.02 | *a* = 5.850(1), *b* = 3.362(1), *c* = 6.292(1) [17] |
| CrAs | *Pnma* | *a* = 5.65204(4), *b* = 3.46493(2), *c* = 6.20866(3) | -- | 0.0286 | 0.0429 | 2.11 | *a* = 5.6490(6), *b* = 3.4609(6), *c* = 6.2084(7) [18] |
| MnAs | *P*6$_3$/*mmc* | *a* = 3.72136(2), *c* = 5.70662(4) | MnO (1.8) | 0.0188 | 0.0250 | 1.36 | *a* = 3.722, *c* = 5.702 [20] |
| FeAs | *Pnma* | *a* = 5.43927(4), *b* = 3.37252(2), *c* = 6.02573(4) | -- | 0.0118 | 0.0175 | 1.77 | *a* = 5.4420(7), *b* = 3.3727(6), *c* = 6.0278(7) [21] |
| CoAs | *Pnma* | *a* = 5.28194(4), *b* = 3.48788(3), *c* = 5.86539(5) | $CoAs_2$ (2.3) | 0.0112 | 0.0157 | 1.43 | *a* = 5.288(1), *b* = 3.492(1), *c* = 5.871(1) [22] |
| NiAs | *P*6$_3$/*mmc* | *a* = 3.61840(1), *c* = 5.03260(3) | $NiAs_2$ (2.4) | 0.0202 | 0.0359 | 3.62 | *a* = 3.618, *c* = 5.032 [19] |
| ZrAs | *P*6$_3$/*mmc* | *a* = 3.80461(1), *c* = 12.86558(4) | $ZrAs_2$ (0.3), $ZrO_2$ (0.7) | 0.0277 | 0.0388 | 2.34 | *a* = 3.80, c = 12.87 [23] |
| NbAs | *I*4$_1$/*md* | *a* = 3.45188(1), *c* = 11.67289(5) | $NbAs_2$ (4.6), $NbO_2$ (0.9) | 0.0291 | 0.0414 | 2.39 | *a* = 3.4520(1), *c* = 11.6754(7) [24] |
| RuAs | *Pnma* | *a* = 5.71685(8), *b* = 3.33790(5), *c* = 6.31294(9) | -- | 0.0292 | 0.0460 | 2.24 | *a* = 5.70(1), *b* = 3.25(1), *c* = 6.27(1) [25] |
| HfAs | *P*6$_3$/*mmc* | *a* = 3.76498(1), *c* = 12.67188(5) | $HfO_2$ (8), ZrO (5), $HfAs_2$ (0.7) | 0.0492 | 0.0684 | 1.09 | *a* = 3.7681(3), *c* = 12.703(2) [26] |
| TaAs | *I*4$_1$/*md* | *a* = 3.43653(1), *c* = 11.6460(1) | $TaAs_2$ (2.1), $Ta_2O_5$ (0.4) | 0.0421 | 0.0562 | 1.82 | *a* = 3.4367(2), *c* = 11.6437(6) [24] |

Figure 3 also shows electrical resistivity data collected in a field of 8 Tesla for those materials which display deviations from the 0 Tesla data. Large positive magnetoresistance (MR), $100(\rho_{8T} - \rho_{0T})/\rho_{0T}$, is observed at 2 K for ScAs (~115%), VAs (~40%), MnAs (~225%), NbAs (~335%), and TaAs (~110%). Large positive MR persists at room temperature for MnAs (~90%), NbAs (~75%), and TaAs (~25%). For NbAs, $\rho_{8T}(T)$ first increases then decreases with temperature. There is no apparent trend in the electrical resistivity data among transition-metal arsenides; this may be partly due to differing crystal structures and valence state of *T*. For example, isostructural VAs, FeAs, CoAs and RuAs show significant differences in ρ(T) behavior. Similar metallic behavior occurs for isostructural and isovalent group 4 materials of TiAs and ZrAs, but no trend holds for group 5 materials NbAs and TaAs.

While a large magnetocaloric effect is well-documented in MnAs [37, 49], this is the first observation of large magnetoresistance on *T* = Sc, V, Nb, Ta, and also on bulk MnAs material. A study aimed at understanding in the latter arsenides should reveal interesting intrinsic and anisotropic results.

Among the binary arsenides, Seebeck coefficient has only been reported for ScAs [28], while thermal conductivity has only been studied for MnAs [50]. Temperature dependence of Seebeck coefficient and thermal conductivity for *T*As is shown in Figure 4. The room temperature values of *S*(T) and κ(T) are summarized in Table 2. The Seebeck coefficient values vary between +40 μV/K and -40 μV/K over the measured temperature range of 2 K to 300 K (Fig. 4a). The values of thermal conductivity are less than 18 W/(m K) in this temperature range.

ScAs and NiAs have small values for the Seebeck coefficient (< -3 μV/K) and change little with temperature. For *T* = Ti, V, Co, Ni, Zr, and Ru, Seebeck coefficients are negative between 2 K and 300 K, suggesting that electrons are the dominant charge carriers. For FeAs, *S* is negative above $T_N$ and is positive below. Such sign changes suggest a competition between multiple bands in FeAs. Anisotropic Hall coefficient ($R_H$) with field along *a*-axis gives $R_H(300K) \approx 1 \times 10^{-4}$ cm$^3$/C, then changes sign close to $T_N$ to $R_H(70K) \approx -3.5\times 10^{-4}$ cm$^3$/C, and finally becomes positive again below 50 K [43]. The presence of both electron and hole charge carriers was inferred for FeAs and also ScAs in band structure calculations [58, 60]. For NbAs, *S* is positive at room temperature, but is negative below ~230 K.

NiAs has the highest thermal conductivity (κ) at room temperature, and also the highest electrical conductivity (σ) suggesting that the electronic contribution to the thermal conductivity ($\kappa_e$) is dominant. Using Wiedemann-Franz law [61], $\kappa_e/\sigma = LT$ with $L = 2.44\times10^{-8}$ WΩ/K$^2$, gives $\kappa_e$(300K)= 14.1 W/(m K) for NiAs that amounts to ≈ 83% of the total thermal conductivity. For semiconducting TaAs, $\kappa_e$(300K)= 0.36 W/(m K), which indicates ≈ 85% lattice contributions to the thermal conductivity. The low-temperature peaks in thermal conductivity data, which is typical for crystalline materials due to umklapp phonon-phonon interactions, are mainly suppressed, likely due to polycrystalline nature of these materials and presence of grain boundaries.

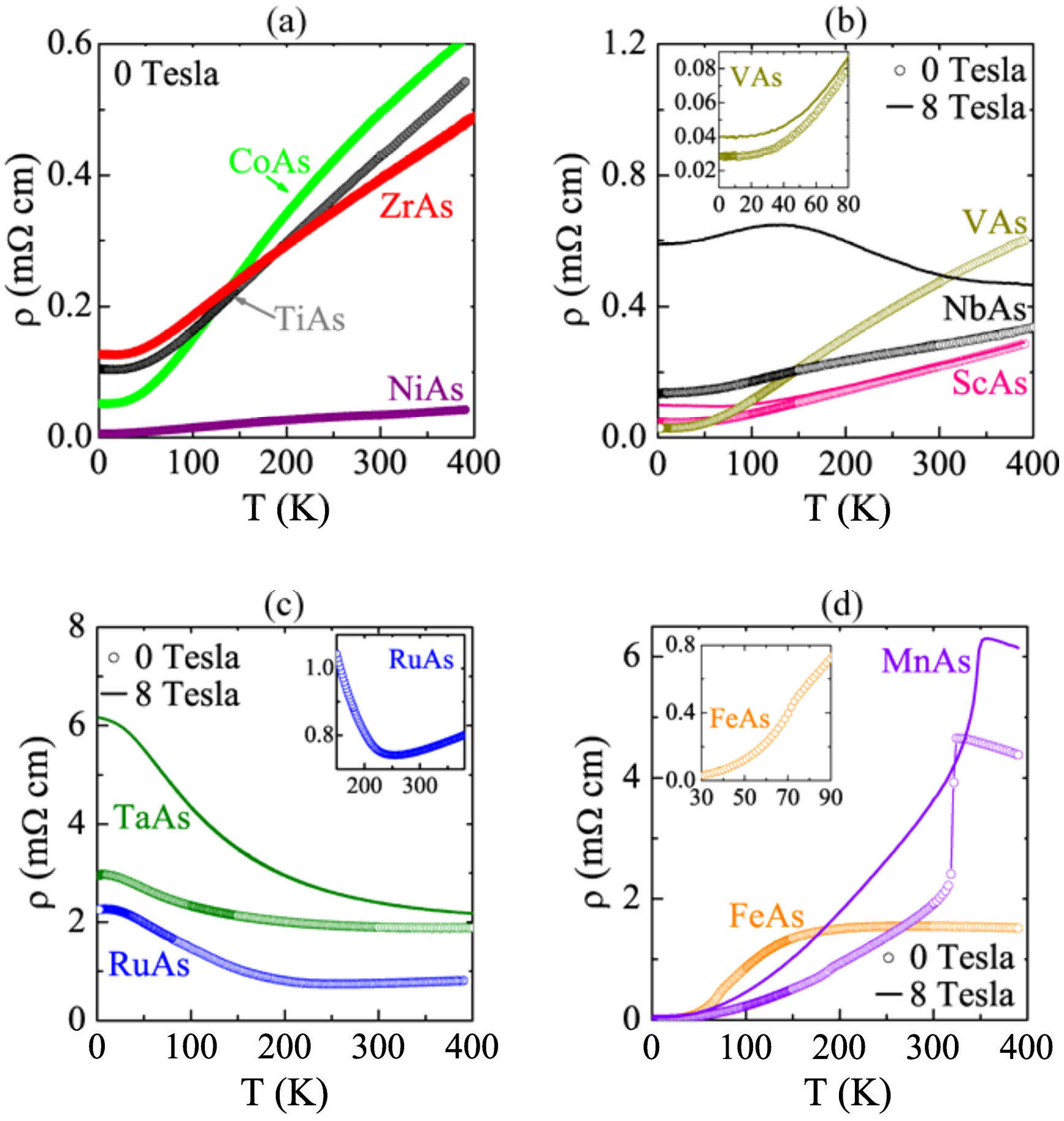

**Figure 3.** Temperature dependence of electrical resistivity, ρ(T), for *T*As with *T* = (a) Ti, Co, Ni, Zr, (b) Sc, V, Nb, (c) Ta, Ru, (d) Mn and Fe. Circles represent 0 Tesla data. Materials displaying magnetoresistance at 8 Tesla are shown with a solid line. Inset in (b) shows magnetic-field effects below 80 K for VAs. Inset in (c) shows the change in slope in ρ(T) for RuAs. Inset in (d) shows the enlarged region in zero-field FeAs resistivity data.

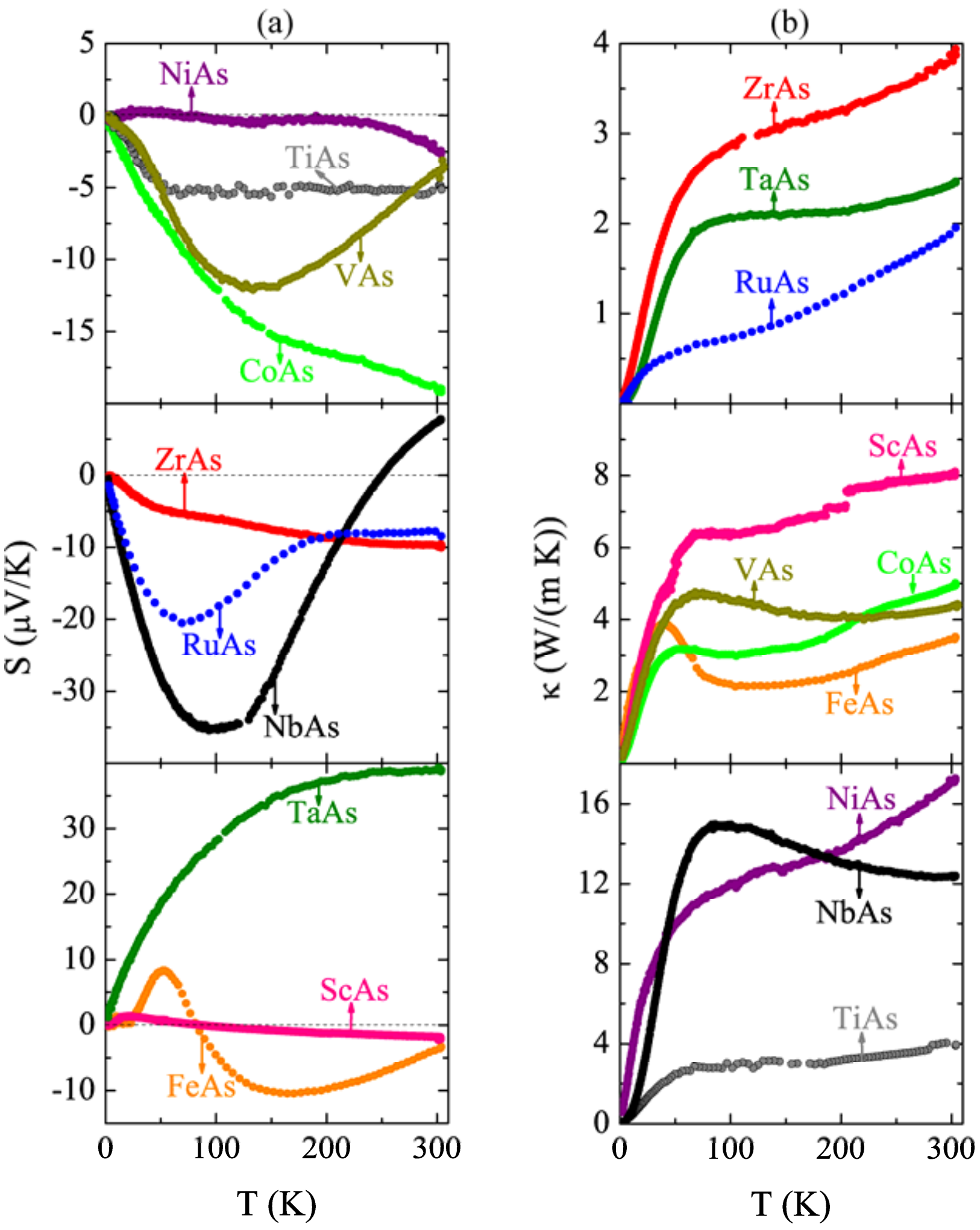

**Figure 4:** For transition metal binaries, the temperature dependence of (a) Seebeck coefficient and (b) thermal conductivity.

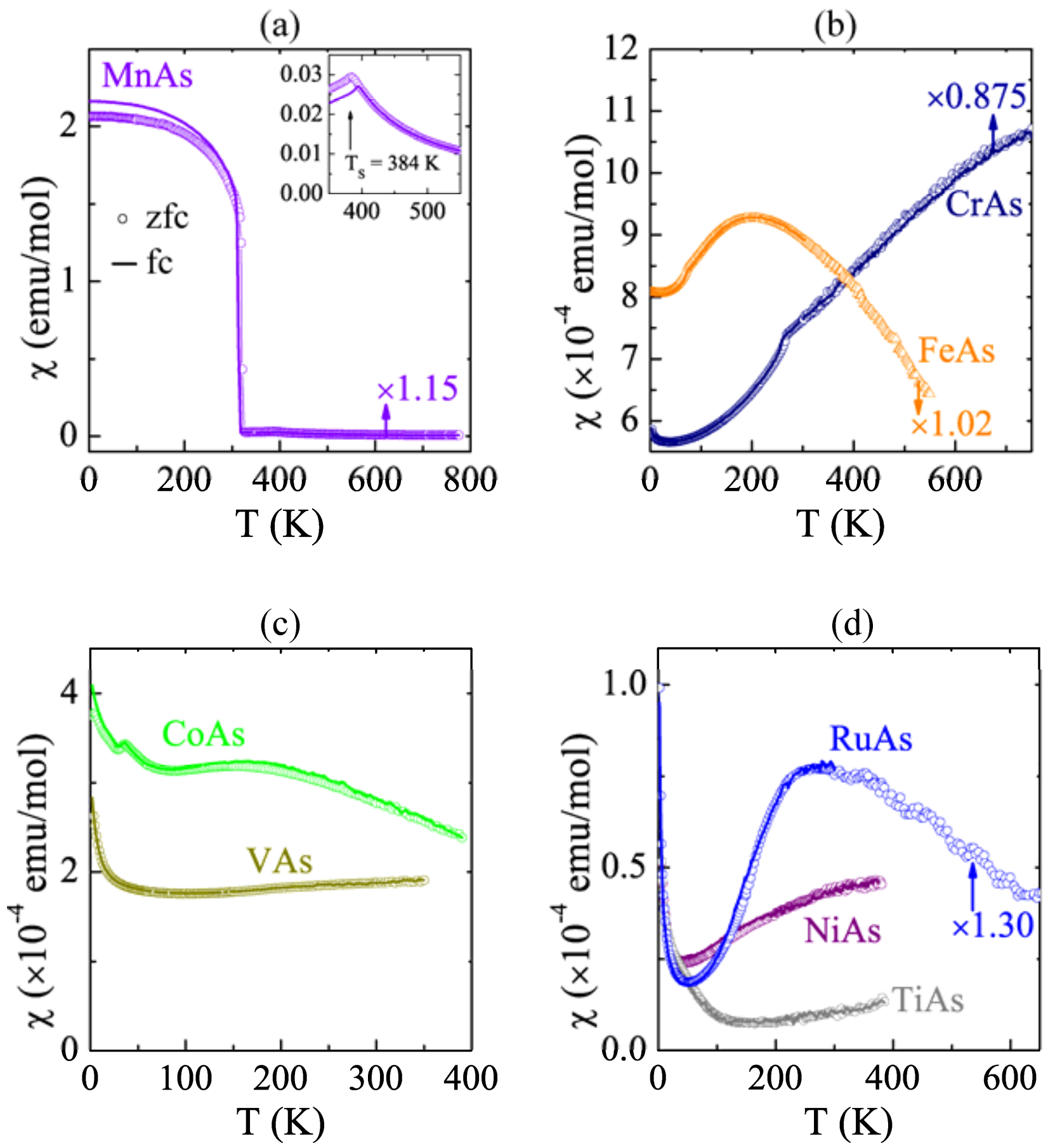

**Figure 5:** Magnetic susceptibility versus temperature $\chi$(T) for *T*As transition-metal binaries. Zero-field cooled (zfc) and field cooled (fc) data are shown as open circles and lines, respectively. A scaling factor for to match high-temperature oven data above ≈ 350 K was required for *T* = Mn, Fe, Cr, and Ru.

Results for the temperature- dependence results for *T*As under 1 Tesla field are shown in Figure 5. It is reported that *T* = Nb and Ta are diamagnetic [48, 49], and in addition to them, we find that ScAs and ZrAs to be diamagnetic as well (data not shown). For the latter two materials, χ increases to positive values below ~ 50 K, which are likely due to paramagnetic impurities detected in x-ray diffraction data (Table 1). For MnAs (Fig. 5a), χ(T) has a strong downturn above $T_C$ = 320 K, a result of already well-characterized ferromagnetic transition that coincides with the structural transition [20, 62, 63]. Above this $T_C$, there is a second structural transition that is driven by chemical-bonding interactions [20, 35]. The sharp small transition at $T_S \approx 384$ K is featured here for the first time in χ(T) and results in divergence of zfc/fc data (inset of Figure 5a). Among the *T*As studied here, MnAs is the only material that fits the Curie-Weiss law according to $\chi(T) = C/(T-\theta) + \chi_0$, with Curie constant (*C*), Weiss temperature (θ), and the temperature-independent term ($\chi_0$), which accounts for Pauli and Van Vleck paramagnetism and also core and Landau diamagnetism. The values thus derived are $\mu_{eff} = 4.26\ \mu_B$ and $\theta \approx 302$ K. For *T*As with *T* = Fe and Cr (Fig. 5b) the χ(T) values are small and less than ~ $10^{-3}$ emu/mol. The data for FeAs (Fig. 5b) matches well with the report on a single crystal [43], featuring a kink at $T_N \approx 70$ K followed by a broad hump, beyond which χ(T) continues to decline modestly. For CrAs, there is a feature at $T_N \approx 260$ K that corresponds to an antiferromagnetic helical ordering [18, 32]. The behavior of χ(T) is very similar to that observed in spin-density-wave systems of Cr metal [64] and $A\text{Fe}_2\text{As}_2$ (*A*= Ba, Sr) [65], which are multiband materials consisting of both electron and hole Fermi surfaces. For *T* = Co and V, χ(T) values are smaller than $4 \times 10^{-4}$ emu/mol. VAs gives approximately temperature-independent Pauli paramagnetic behavior as reported in literature [49] with a small upturn in χ below 25 K, while CoAs shows slightly temperature-independence behavior with a small kink at ~ 36 K and a broad hump around 150 K (Fig. 5c). The smallest magnetic susceptibility is found for *T* = Ti, Ni, and Ru with values less than $8 \times 10^{-5}$ emu/mol. They all have similar decrease in χ at low temperatures. TiAs shows temperature-independent behavior between 120 K and 400 K, NiAs increases with temperature, RuAs shows a steep increase above ~ 50 K followed by a broad maximum between ~ 200 K to 300 K region, with a decrease in χ above. In a recent report, RuAs was described as a Pauli paramagnet for which the broad peak in χ(T) is attributed to a pseudo-gap formation [47].

The field-dependent magnetization results for CrAs, MnAs, and RuAs are shown in Figure 6. For MnAs, the saturation moment of $\mu_{Sat} \approx 3.6\ \mu_B$ at 2 K is in a good agreement with the values of 3.3 $\mu_B$ at 4 K determined from neutron diffraction experiments [20]. Assuming the spin only value, the calculated effective moment from the fit corresponds to 3.4 unpaired electrons (*n*) per Mn according to $\mu_{eff} = [n(n+2)]^{1/2}$. For CrAs, M(H) dependence is linear at 2 K and 295 K (Figure 6b), consistent with that already reported [34] and similar to the data for NiAs (not shown). M(H) dependence is also linear for FeAs and TiAs at 2 K and 295 K, although there is a small curvature below 0.5 T at 2 K (data not shown). For RuAs, M(H) data is linear at room temperature, but not 2 K (Fig. 6c).

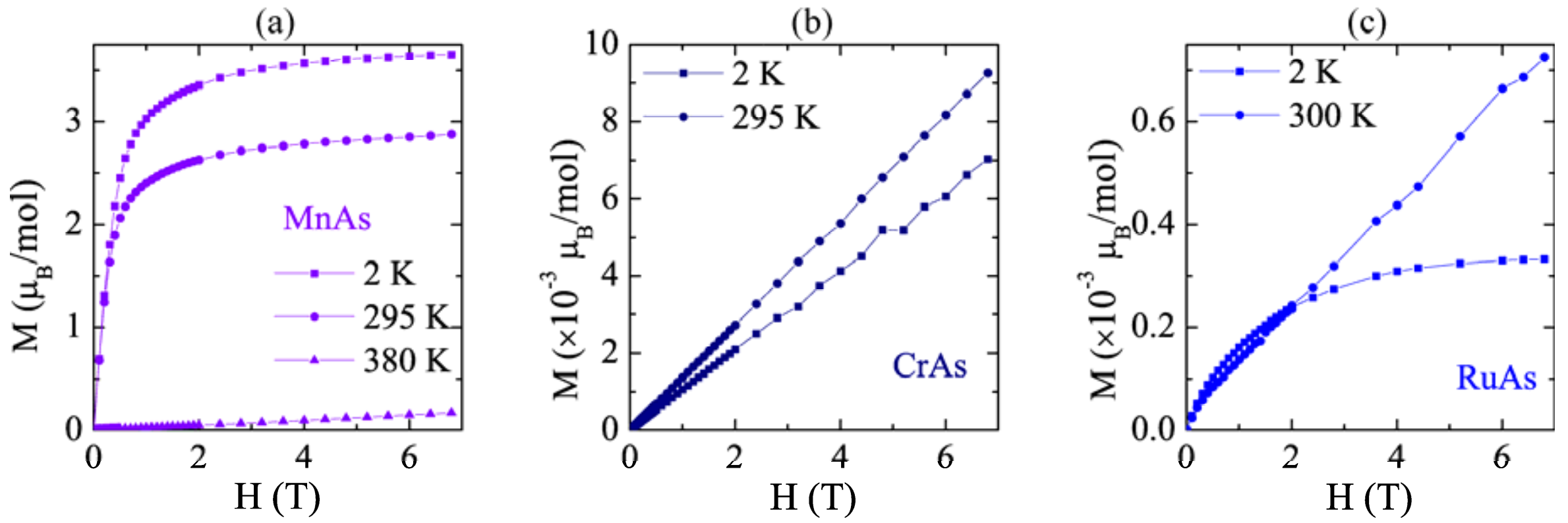

**Figure 6:** Field dependent magnetization curves for *T*As (*T* = Mn, Cr and Ru) at various temperatures.

**Table 2**: For *T*As binary arsenides, the room-temperature resistivity (ρ), Seebeck coefficient (*S*) and thermal conductivity (κ) values.

| | $\rho$ (mΩ cm) | $S$ (μV/K) | $\kappa$ (W/(m K)) |
|---|---|---|---|
| ScAs | 0.219 | -2.05 | 8.02 |
| TiAs | 0.430 | -5.14 | 3.93 |
| VAs | 0.473 | -3.61 | 4.38 |
| MnAs | 1.932 | - | - |
| FeAs | 1.543 | -3.36 | 3.51 |
| CoAs | 0.496 | -18.97 | 4.96 |
| NiAs | 0.034 | -2.57 | 17.18 |
| ZrAs | 0.393 | -9.76 | 3.88 |
| NbAs | 0.281 | 7.63 | 12.37 |
| RuAs | 0.754 | -7.81 | 1.88 |
| TaAs | 1.893 | 38.94 | 2.43 |

Temperature dependent heat capacity data in the form of $C_p(T)$ and $C_pT^{-1}(T^2)$ are shown in Figure 7. FeAs shows an anomaly at $T_N \sim 66$ K that corresponds to the antiferromagnetic onset (Fig. 7a). This transition has been carefully studied in literature, though the exact transition temperature seems to be sample-dependent and literature values vary from 70 K to 77 K [21, 43-45, 66]. For the remaining *T*As with *T* = Sc, Ti, V, Mn, Fe, Co, Ni, Zr, Nb, and Ta, we found no evidence for phase transitions in heat capacity in the 1.9 to 200 K range. Sommerfeld electronic coefficients ($\gamma$) were extracted from low temperatures (below 8 K) from fits to $C_p/T = \gamma + \beta T^2$. The $\gamma$ values range from 0.28 mJ/(K$^2$mol) for NbAs to 7.97 mJ/(K$^2$mol) for MnAs. Debye temperatures ($\theta_D$) were estimated by performing fits to data in the range of 1.9 K to 200 K using the Debye model of $C_V(T) = 9R\left(\frac{T}{\theta_D}\right)^3 \int_0^{\theta_D/T} \frac{x^4 e^x}{(e^x-1)^2} dx$, where $C_v$ is the Debye lattice heat capacity due to acoustic phonons at constant volume ($V$). These fits were further improved [67] by considering the sum of contributions from $C_v$(T) and predetermined electronic heat capacity $\gamma T$ according to $C_p(T) = \gamma T + n\, C_v(T)$, where $n$ is the number of atoms per formula unit. In this manner, the $\theta_D$ values were found to be in the range of 276 K to 399 K. For example, $\theta_D$ values for MnAs, FeAs and NbAs are respectively 276 K, 355 K and 399 K. These values compare well with previously reported $\theta_D$ =310 K for MnAs [39], $\theta_D$ =353 K for FeAs [44] and $\theta_D$ = 370 K for CrAs [35].

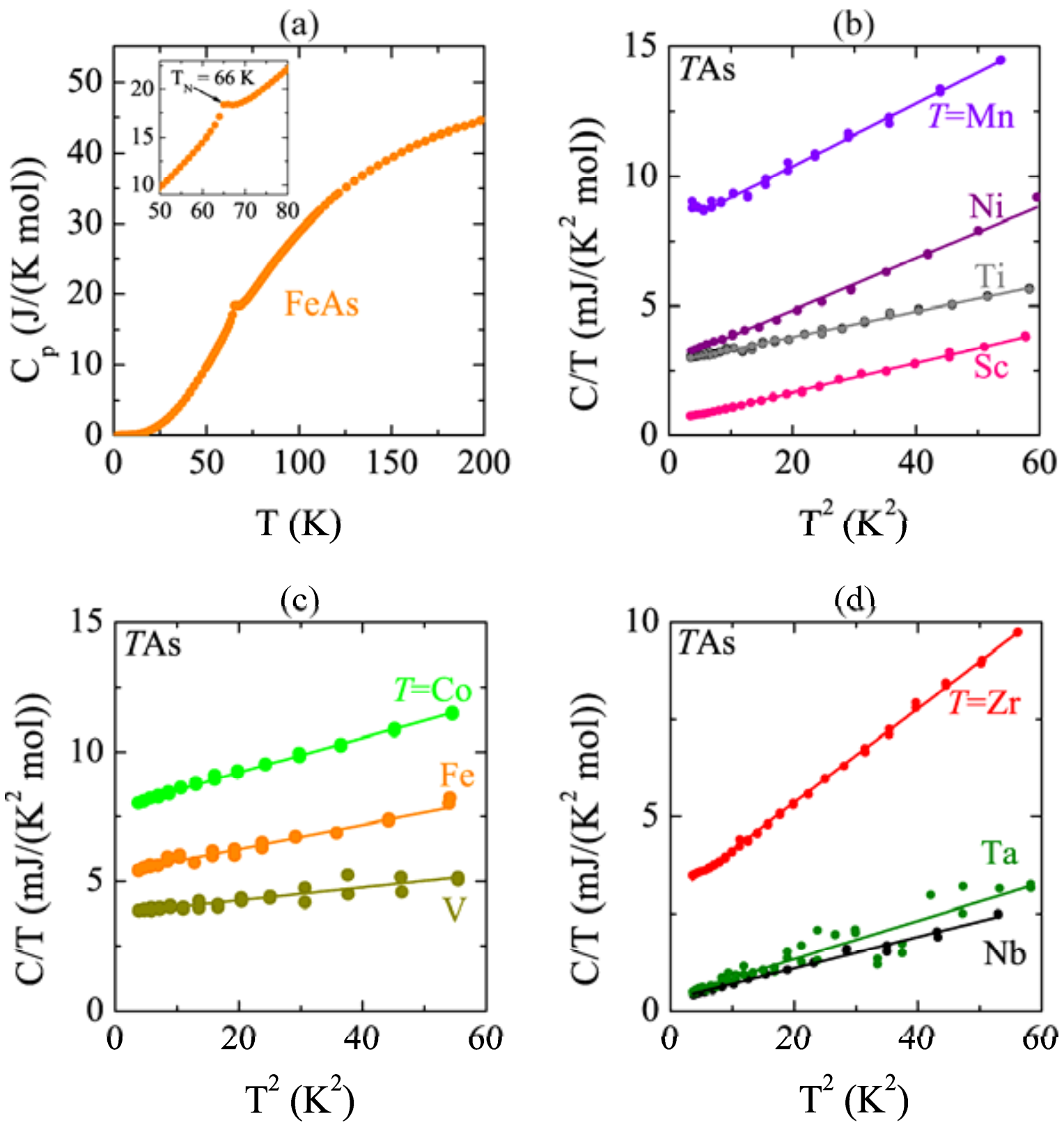

**Figure 7:** Heat capacity versus temperature in form of (a) $C_p(T)$ for FeAs, and (b, c, d) $C/T(T^2)$. Plots of The data are the filled circles, and the lines are the linear fits to the low-temperature data.

## Conclusions

There has not been empirical evidence relating the properties of elements and the superconducting compounds they form. For example, neither the metallic copper (non-magnet) nor iron (ferromagnetic) superconduct, but each is a crucial component to one of the two classes of Fe-based and cuprate high-temperature superconductors. Similarly, we find no simple association between the behaviors of *T*As binaries and the arsenic-based superconductors which dominate the FeSC family.

The most studied transition-metal arsenides are MnAs and FeAs, according to our literature survey. Interest in understanding the complex magnetism and magnetocaloric effect in MnAs has spanned decades, whereas the renewed interest in FeAs is due to the discovery of iron-arsenide based superconductors. We report synthesis and general thermodynamic and transport properties of *T*As with *T* = Sc, Ti, V, Cr, Mn, Fe, Co, Ni (3*d*), Zr, Nb, Ru (4*d*) and Ta (5*d*). The room-temperature structural lattice parameters and values for resistivity, Seebeck coefficient, and thermal conductivity are summarized in Tables 1 and 2. All binaries show metallic behavior except *T* = Ta and Ru. Arsenides of *T* = Ta, Nb, Sc and Zr are diamagnetic, while T = Co, V, Ti, Ni and Ru are approximately Pauli paramagnetic. Binaries of *T* = Fe and Cr have long-range antiferromagnetic order below $T_N \approx 71$ K and $T_N \approx 260$ K, respectively. MnAs is a ferromagnetic below $T_C \approx 317$ K and it undergoes hexagonal-orthorhombic-hexagonal transitions at $T_S \approx 317$ K and 384 K, respectively. For *T*As, Seebeck coefficients vary between + 40 μV/K and - 40 μV/K in 2 K to 300 K range, whereas thermal conductivity values stay below 18 W/(m K). The low values of Seebeck coefficients are expected considering the temperature dependence of electrical resistivity results, which prevent them from being useful thermoelectric materials. We find significant magnetoresistance behavior in *T* = Sc, V, Mn, Nb, and Ta binaries; a study aimed at understanding magnetoresistance in the latter arsenides should reveal interesting intrinsic and anisotropic results.

Relating the results of this study to the iron-arsenide based superconductivity, we conclude that CrAs may have the most similarity to FeAs in terms of crystal structure (MnP-type) and magnetism (antiferromagnetic helical ordering). It is interesting to note that Rh-doped RuAs superconductors [47] also adopt MnP-type structure making binary arsenides, with this structure-type, candidates for chemical-doping studies. Because RuAs is approximately Pauli-paramagnetic, as are VAs and CoAs, they may be the best candidates to explore novel superconductivity. This manuscript may serve as a reference for effects of binary *T*As impurities caused by transition-metal doping of iron-based arsenides.

## Acknowledgments

This work was supported by the Department of Energy, Basic Energy Sciences, Materials Sciences and Engineering Division. We acknowledge A. F. May for his assistance with heat capacity data analysis.